\documentclass[a4paper]{article}

\usepackage{spconf}
\usepackage{graphicx}
\usepackage{MnSymbol,wasysym}
\usepackage{tikz}
\usepackage{siunitx}
\usepackage{todonotes}
\usepackage{url}

\usepackage{xspace}

\usepackage[minnames=2,maxnames=3,style=ieee,doi=false,url=false,isbn=false,backend=bibtex]{biblatex}
\usepackage[unicode=true,pdfusetitle,%
 pdfauthor={Jacob J Webber, Cassia Valentini-Botinhao, Evelyn Williams, Gustav Eje Henter, Simon King},%
 pdfkeywords={speech synthesis, neural vocoder, differentiable DSP, representation learning},%
 bookmarks=true,bookmarksnumbered=true,bookmarksopen=false,%
 breaklinks=true,pdfborder={0 0 0},backref=false,colorlinks=true,citecolor=blue]{hyperref}

\def\autovocoder{\textit{autovocoder}\xspace}
\def\Autovocoder{\textit{Autovocoder}\xspace}

\def\F0{$f_0$\xspace}

\title{Autovocoder: Fast Waveform Generation from a Learned Speech Representation using Differentiable Digital Signal Processing}

\name{Jacob J Webber$^1$, Cassia Valentini-Botinhao$^1$, Evelyn Williams$^1$, Gustav Eje Henter$^2$, Simon King$^1$\thanks{JJW is funded by the Engineering and Physical Sciences Research Council Centre for Doctoral Training in Pervasive Parallelism (grant EP/L01503X/1) at the University of Edinburgh, School of Informatics. GEH is partially supported by the Wallenberg AI, Autonomous Systems and Software Program (WASP) funded by the Knut and Alice Wallenberg Foundation.}}
\address{
  $^1$The Centre for Speech Technology Research, University of Edinburgh, United Kingdom\\
  $^2$Division of Speech, Music, and Hearing, KTH Royal Institute of Technology, Stockholm, Sweden}
\begin{document}

\maketitle

\begin{abstract}
% The dominance of the mel-spectrogram as an intermediate representation in text-to-speech systems is being threatened by end-to-end systems and by learned intermediate representations like Wav2Vec 2.0.

Most state-of-the-art Text-to-Speech systems use the mel-spectrogram as an intermediate representation, to decompose the task into acoustic modelling and waveform generation.

A mel-spectrogram is extracted from the waveform by a simple, fast DSP operation, but generating a high-quality waveform from a mel-spectrogram requires computationally expensive machine learning: a neural vocoder.
Our proposed ``autovocoder'' reverses this arrangement. We use machine learning to obtain a representation that replaces the mel-spectrogram, and that can be inverted back to a waveform using simple, fast operations including a differentiable implementation of the inverse STFT.

The autovocoder generates a waveform 5 times faster than the DSP-based Griffin-Lim algorithm, and 14 times faster than the neural vocoder HiFi-GAN. We provide perceptual listening test results to confirm that the speech is of comparable quality to HiFi-GAN in the copy synthesis task.
%Our approach demonstrates how differentiable waveform generation can facilitate the creation of fast and natural end-to-end speech-synthesis systems for the future.
\end{abstract}
\noindent\textbf{Index Terms}: speech synthesis, neural vocoder, differentiable DSP, representation learning

\section{Introduction}

%The task of speech synthesis is to convert text, frequently converted into a sequence of phones, into a waveform. The nature of waveforms makes them challenging to generate. Sequence-to-sequence \textit{acoustic models} have been extremely successful at efficiently mapping from phones to an audio representation, with the task of synthesizing a waveform from that generally done separately by a \textit{waveform synthesizer}. Frame-based audio representations, such as mel-spectrograms, are widely used as the intermediate representations between these two constituent systems. The frame rate of the audio representation is much closer to the rate of phonemes than to waveform sample rate, and the audio representation is lossy with respect to the waveform. This simplifies the task for the acoustic model but creates a challenge for the vocoder.
Generating a natural-sounding speech waveform is a challenging task.
For this reason, text-to-speech (TTS) is usually divided into \emph{acoustic modelling} to map input text to an intermediate acoustic representation, which is then input to a \emph{waveform generator}.
The acoustic representation is, -- in almost all state-of-the-art systems -- a mel-spectrogram. The frame rate of this spectrogram is much lower than the waveform sampling rate, which simplifies the task for the acoustic model but creates a challenge for the waveform generator. Other tasks, such as Voice Conversion (VC), also involve the generation of a waveform from a spectrogram.

%Waveform synthesizers achieved extremely natural-sounding results with the advent of deep-learning models starting with WaveNet \cite{wavenetvocoder}. However, because many of these systems are autoregressive (mirroring the autocorrelated nature of speech waveforms), they require intensive computation.
%This is expensive, energy-intensive, and precludes on-device synthesis.

Beginning with WaveNet \cite{wavenetvocoder}, deep learning has achieved extremely natural-sounding waveform generation, but many such \textit{neural vocoders} are autoregressive (to model the autocorrelated nature of speech waveforms) which is computationally expensive, not amenable to parallelisation, and precludes on-device synthesis.
More recent neural vocoders reduce computational cost either by leveraging knowledge of speech production, for example with the use of traditional signal-processing ideas in Neural Source-Filter (NSF) models \cite{nsf} and LPCNet \cite{lpcnet}, or by making use of generative modelling to remove the autoregression, as in HiFi-GAN \cite{hifigan}.

Our proposed system, \autovocoder, generates high-quality waveforms very efficiently from a frame-based representation of similar size to the mel-spectrogram used in a typical TTS or VC system.
The system uses machine learning to define that acoustic representation, then leverages fast, differentiable DSP (DDSP) operations for decoding.
This reverses the conventional arrangement, where fast signal processing is used for the one-off task of encoding waveforms into acoustic representations (mel-spectrograms), forcing deployed systems to use slow, difficult to parallelise, and computationally expensive models to generate from these features every time a waveform is synthesised: Table \ref{tab:dsp_ml}.

\begin{table}[!t]
\footnotesize
    \caption{Comparing encoding and decoding of the proposed system with the current state-of-the-art (SOTA).}
\vspace{2mm}
    \centering
    \begin{tabular}{c|p{2.5cm}p{2.5cm}}
         System & Acoustic feature extraction (encoder) & Waveform generation (decoder)  \\ \hline
         \rule{0pt}{2ex}

         SOTA & fast DSP  & slow neural network \vspace{1mm} \\ 
         \rule{0pt}{2ex}
         \Autovocoder & fast DDSP + simple neural network & simple neural network + fast DDSP
    \end{tabular}
    \label{tab:dsp_ml}
\vspace{-5mm}
\end{table}

\Autovocoder combines the respective strengths of DSP and deep learning. DSP is fast and efficient, while machine learning can yield rich and maximally informative representations (for a given dimensionality).
The mel-spectrogram only represents spectral magnitudes and discards phase, even though this is known to be beneficial for both speech recognition \cite{Loweimi20216738} and speech synthesis \cite{magphase}. The learned representation of \autovocoder is not required to discard phase.  
%
%Machine-learning can yield a rich and maximally-informative (for a given dimensionality) representation of the waveform.
%By learning the acoustic representation, we can mitigate redundancies and information loss present in conventional representations from signal processing.
%, if an appropriate phase representation is used.
For signal generation, we use the inverse discrete short-time Fourier transform (iSTFT) and overlap-add, which can be executed extremely efficiently on modern hardware.
%Mel-spectrogram features, for example, completely exclude phase, even though phase been found to contain important information for speech processing \cite{Loweimi20216738} and can improve speech-synthesis models, if represented appropriately \cite{magphase}.
%Machine learning can avoid this loss of salient information and represent phase in a useful form.
%We couple our learned representations with the inverse discrete short-time Fourier transform (iSTFT) and overlap-add for signal generation, which can be executed extremely efficiently on modern hardware.
Experimental results show that the proposed approach offers high-quality speech waveform generation whilst also being extremely fast.

\section{Background \& related work}

\subsection{Parallelism and autoregression}

The rise of machine learning for image, speech, and language processing has been enabled by rapid advances in parallel computing performance. Most modern machine learning is highly parallelisable, but
%Deep learning can largely be reduced to a sequence of matrix multiplications, which can be executed extremely efficiently on modern parallel hardware such as GPUs and TPUs. However, some kinds of machine learning algorithm actually incur performance penalties on such hardware.
algorithms parallelise to different degrees.
%Some sequence models are \emph{autoregressive}, meaning that the synthesis of each step in a sequence requires the value of prior steps to be computed first, not in parallel. 
Autoregressive sequence models require values from prior steps in the computation: output must be computed sequentially and not in parallel.
Conversely, \emph{embarrassingly} parallel algorithms require no communication between decomposable computations. Many recent TTS acoustic models are not autoregressive \cite{fastpitch, fastspeech}; they are fast because a sequence can be processed in parallel without depending on prior steps. 
% This allows all computations in a long sequence to be distributed across many processing units and processed in unison.
%
%The non-autoregressive TTS models still output a mel-spectrogram and therefore a vocoder is required. 
% Each of the steps in these sequences is fairly self-contained, containing information specific to the time window that it represents. 
Non-autoregressive waveform synthesis has proved more elusive.
% , because the human auditory system does not perceive the individual samples in a waveform, but rather groups of frequencies, similar to those represented by the mel spectrogram.
%These frequencies cannot be derived from individual samples, but only by large windows of samples.
Autoregression is a straightforward simple approach for modelling between-sample dependencies, and was used by the first machine-learned waveform generator, WaveNet \cite{vandenoord2016wavenet,tamamori2017speaker}.
%WaveNet \cite{wavenetvocoder} was the first attempt at machine-learned waveform synthesizer.
WaveNet suffered very significant performance issues, which were to some extent addressed by Parallel WaveNet \cite{parallel_wavenet}. More recent systems like WaveRNN \cite{wavernn} have achieved better computational performance through more strategic use of autoregressive elements and careful engineering. Even more recently, sophisticated generative models such as GANs have enabled parallel waveform synthesis \cite{hifigan}. However, these systems still rely on learned methods to generate highly correlated signal samples, which is inefficient with respect to both training data and compute.

\subsection{Vocoding based on explicit harmonic synthesis}

%Some recent systems have mitigated some of the challenges of generating autocorrelated waveforms by explicitly generating periodic content.
Some recent systems have considered generating waveforms with appropriate correlations by explicitly generating the periodic content.
The Neural Source-Filter (NSF) model \cite{nsf}, employs a series of sinewave generators that generate harmonics. This requires a known input \F0.
Artefacts for speech with less harmonicity in the source (e.g., breathy voice) were addressed in \cite{nsf_cyclic} by using cyclic noise rather than sinewaves for the source.
%The original system generated artifacts for source sounds which are less harmonic (for example breathy or noisy voices), which was addressed in \cite{nsf_cyclic} by using cyclic noise as the source rather than sinewaves.
Another important example of DSP-informed waveform synthesis for TTS is LPCNet \cite{lpcnet}, which employs an RNN to drive a waveform generator based on Linear Predictive Coding. The speech signal is encoded into 20 DSP-derived parameters at a frame rate of \SI{10}{\milli \second}. 
% This has since been used to generate speech alongside an LSTM-based speech syntheser in \cite{lpcnet2}.
% 
% Their assumptions about signals containing only one pitched source could limit flexibility, as this is built into the construction of the model.
% 
%These systems trace a lineage back to traditional DSP-based source-filter vocoders \cite{dsp_source_filter}, notably STRAIGHT \cite{straight}.
% 
Both NSF and LPCNet differ from the work in this paper by continuing to depend on DSP-derived features.

While iSTFTNet \cite{istftnet}, developed in parallel with our work, also uses the iSTFT for synthesis, it has key differences in that it employs a much smaller FFT and hop size (16 vs.\ 1024 and 4 vs.\ 256 respectively), which is likely to limit computational performance. These settings may be necessary because it still depends on the mel-spectrogram. Puffin \cite{puffin23}, another vocoder developed in parallel to ours, relies on a pitch synchronous iSTFT operation to support low-complexity higher sampling-rate generation.

Recent work, such as wav2vec 2.0 \cite{wav2vec2}, has used self-supervised learned audio representations. These have been used in TTS \cite{vqtts, wavthruvec}, but they necessitate a \emph{more} complex waveform decoder, plus \F0 input. Our representations are designed to be efficiently invertible.
% \todo{Jacob just added this}

\Autovocoder is trained as an autoencoder. Other systems have used autoencoders for unsupervised speech feature learning \cite{unsupervised}; however, they emphasise the use of the features for other tasks, and are not designed to perform very fast decoding back to a waveform. Unlike neural speech codecs like \cite{soundstream}, we do not target extremely low bit rates, but rather a frame-based representation that could directly replace the mel-spectrogram in applications including TTS.

%\todo{I need to mention/cite WavThruVec and VQTTS, which use learned representations for TTS}
%Autoencoders are deep learning networks that derive representations of signals using \emph{unsupervised learning}. The aim of an autoencoder is to learn to represent a signal or input with a reduced dimensionality. This results in compressed representations. This is achieve by making assumptions based on knowledge of the domain. For example, an autoencoder trained on speech will know to generate harmonic structure without each harmonic necessarily being encoded.

\section{The proposed system}

%This work proposes replacing mel-spectrograms as acoustic representations with a learned embedding that has similar frequency and dimensionality. This embedding aims to fulfill the same task as mel-spectrograms do with typical SOTA neural vocoders. 

%Our proposed system resembles an \emph{autoencoder}, in that it simultaneously learns a compressed encoding of a signal, and learns to reconstruct the signal from its compressed representation.
%a deep learning system which encodes and decodes a signal, and in the process derives a compressed representation of the original input. 
\Autovocoder is an autoencoder trained on speech waveforms. The aim is to learn a representation that can replace the conventional, signal processing-based mel-spectrogram.
%aims to replace the signal processing-based mel-spectrogram with a learned representation.
In the current work, we use a dimensionality and frame rate that is comparable to a typical mel-spectrogram, for fair comparison in our experiments.
%the experimental results reported in this paper.
% intended are therefore largely compatible with existing architectures of seq2seq TTS models.
Our focus in this paper is on an architecture whose learned representation can be decoded back to a waveform with very low computational cost.

\subsection{Encoder and decoder}

The encoder converts from time domain to frequency domain using a differentiable implementation of the STFT. The resulting complex spectrum is then used to derive four spectral components: magnitude, phase, real and imaginary.
This use of redundant components is discussed in Section \ref{sec:model_train}.
%The former and latter pairs of these are of course mutually redundant, with each pair jointly being a complete representation of the original signal. The use of redundant spectral components is discussed in Section \ref{sec:model_train}.   
These four spectral components are stacked, with each treated as a channel, and fed into a purely convolutional residual network made up of what we term \emph{basic blocks}. Each such block consists of two 2D convolutional layers of width 3, followed by a 2D batch norm and a ReLU nonlinearity. This basic block also applies a residual, by summing the input with the output, but only if the number of input channels is the same as the number of output channels.

The residual net used in the current \autovocoder architecture consists of 11 basic blocks, the first five having 4 input and output channels, the middle one having 4 input channels but 1 output channel, with the remaining 5 blocks having 1 channel in and out.
That single channel output is fed into a single linear layer that reduces the dimensionality per timestep from $(\mathrm{window size})/2 + 1$ to our \emph{representation size}.
%, which we initially chose to be 128.
%For our representation size, the value of 128 was chosen initially.
This is the dimensionality of a single frame of the learned representation, which we chose to be similar to the frequency dimension of a typical mel-spectrogram in the experiments.

\begin{figure}
    \centering
    \includegraphics[width=\linewidth]{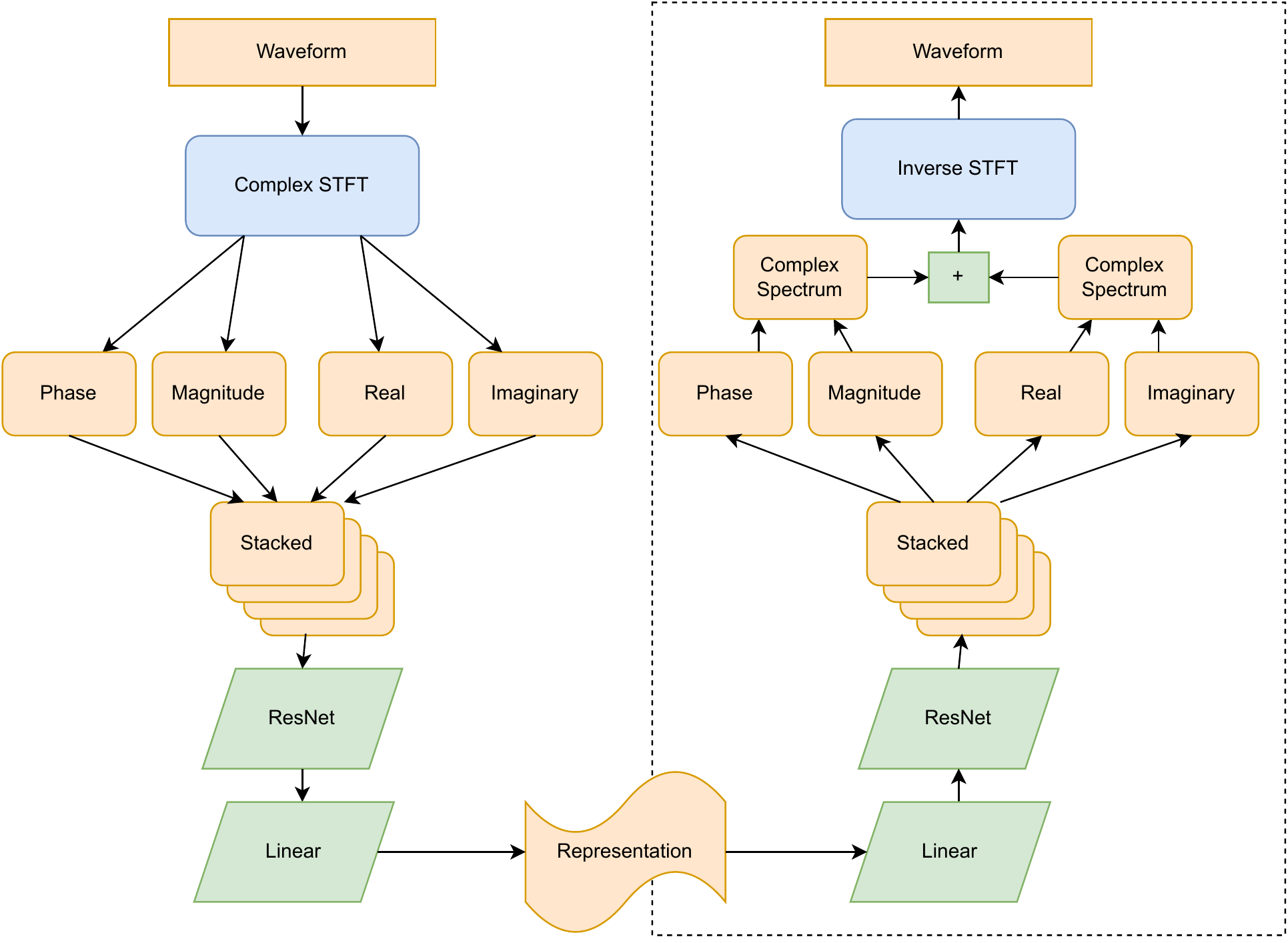}
    \caption{Autovocoder architecture. Dashed box shows decoder}
    \label{fig:autovocoder}
\end{figure}

%\subsection{Decoder}
%The decoder is constructed similarly, with each of the components applied in reverse order making the decoder architecture a mirror of the encoder architecture, with similar hyperparameters.
Our decoder architecture is a mirror of the encoder, with the components applied in reverse order and with similar hyperparameters.
The full architecture is shown in Fig.\ \ref{fig:autovocoder}.
Note that the system has no autoregression at
%frame, sample, or any other
any level. All frames are processed at once by the network, and the subsequent overlap-add to assemble the complete waveform is performed by the differentiable iSTFT function of PyTorch \cite{pytorch}.

\subsection{Training regime}
\label{sec:model_train}

The model is trained as a denoising autoencoder \cite{denoising_autoencoder}. Dropout is used on the embedding during training to increase decoder robustness, with a view to subsequent applications where these features would be generated by a TTS acoustic model. \Autovocoder training adopts the losses from HiFi-GAN \cite{hifigan}: a mel-spectrogram loss and two adversarial losses from multi-scale and multi-period discriminators. In addition, a time-domain loss (per-sample squared error) is added, to improve phase reconstruction.
If the weight of this term is too low, highly audible phase artefacts result.
%We experimented with different weightings for the time-domain loss. Low values resulted in highly audible phase artefacts.

%Our system used an STFT window size of 1024 and a representation size of 128, 192 or 256.
%Unlike work such as Soundstream \cite{soundstream}, we don't target a low bitrate.

% Choosing an appropriate representation size is likely to involve a trade-off in the complexities of the acoustic model and the vocoder.

\subsection{Redundant representations of complex numbers}

Providing redundant representations of the complex spectrogram to the encoder allows the model to learn how best to represent magnitude and phase. For example, \autovocoder could potentially learn that, given a magnitude spectrogram, phase can be very efficiently represented  \cite{s_spect_1}.

For the decoder, three representations of the complex spectrogram were considered. 
For Cartesian and polar, the network generates two output channels: real and imaginary, or magnitude and phase, respectively. 
%parts of the signal, respectively. 
In a third method the network generates all four output channels, then the Cartesian mean of both complex forms is taken. 
Inspecting the ratio between the magnitudes of the polar and Cartesian outputs during autoencoding 
% showed that 
% the polar form had a relatively larger magnitude when the time-domain loss weight was low.
%In this latter arrangement, we measured the ratio of magnitudes of the polar and Cartesian outputs. This ratio was weighted towards the polar form when the time-domain loss factor was low, and was weighted towards the Cartesian form when the time-domain loss was low. Given the reduced phase artefacts in the systems with higher weighted time-domain losses,
% This 
suggested that 
phase is more easily modelled in Cartesian form. Training with only polar output yielded poorer sound quality, as did the third method using all four output channels.

%%%%%%%%%%%%%%%%%%%%%%%%%%%%%%%%%%%%%%%%%%%%%%%%%%%%%%%%%%%%%%%%%%%%%%%%%%%%%%%%%%%%%%%%%%%%%%%%%%%%%%%%%%%5

%%%%%%%%%%%%%%%%%%%%%%%%%%%%%%%%%%%%%%%%%%%%%%%%%%%%%%%%%%%%%%%%%%%%%%%%%%%%%%%%%%%%%%%%%%%%%%%%%%%%%%%%%%%%%

\section{Evaluation}
We compared the performance of \autovocoder with two other waveform generators -- Griffin-Lim and HiFi-GAN -- in output quality and computational performance. 

Griffin-Lim \cite{griflim} is an iterative algorithm for phase recovery that converts magnitude spectrograms into phase-coherent time-domain signals. We used the librosa \cite{librosa}  implementation of Griffin-Lim, which uses 32 iterations.
This was used with the same frame rate (i.e. hop size) and window duration as the STFT used by \autovocoder. Griffin-Lim uses full resolution ground truth magnitude spectrograms (513 bins), without mel-scale dimensionality reduction.

An open source implementation of HiFi-GAN\footnote{\url{https://github.com/jik876/hifi-gan}} was used with a pretrained checkpoint that was trained on the same dataset as \autovocoder using the the \emph{V1} and \emph{V3} generators. The \emph{V1} generator is designed to achieve maximum output quality, and the \emph{V3} generator is designed to run much faster, at the expense of some output quality.

The configuration of \autovocoder evaluated here was determined using a combination of informal listening and the loss on a validation set. The STFT and iSTFT hop size was 256 samples (\SI{12}{\milli\second}) with a window size of 1024 samples (\SI{46}{\milli\second}). 

Unlike a neural speech codec such as \cite{soundstream}, there is no intrinsic need for a small representation size in \autovocoder, since we are not aiming for a low bitrate. Nevertheless, here we use modest representation dimensionalities of 128, 192 and 256, chosen to be comparable to the internal representations in current TTS models \cite{fastpitch, vits}, with a view to a future TTS application of \autovocoder.

\Autovocoder and HiFi-GAN were trained on the single-speaker LJ~Speech \cite{ljspeech17} dataset with waveforms at a sample rate of \SI{22.05}{\kilo\hertz}. We used a pretrained model for HiFi-GAN that had been trained to 2.3 million steps. \Autovocoder was trained with a dropout factor of 10\% and the Adam optimiser for approximately 1 million steps (limited by available compute). 
Evaluations used a test section of the dataset, unseen during training, and identical for HiFi-GAN and \autovocoder.

In this paper, we evaluate copy synthesis. In the case of \autovocoder this means a complete pass through both the encoder and decoder parts of the system. For the comparison systems, it means synthesising from ground truth spectral features extracted from waveforms using DSP.
% This task is less challenging than generating waveforms from synthesised audio.

% \begin{figure}
%     \centering
%     \includegraphics[width=\linewidth]{LaTeX/figures/spec_fig.png}
%     \caption{Spectrogram of original signal and of signal reconstructed after passing through the \autovocoder encoder and decoder}
%     \label{fig:spects}
% \end{figure}

% CASSIA READ UNTIL HERE

\subsection{Listening test}

The system was compared to relevant baselines using a MUSHRA \cite{itu2015method} style evaluation. Each MUSHRA screen presented 7 stimuli to the listener for evaluation. These were \autovocoder (\emph{AV 128}, \emph{AV 192}, \emph{AV 256}), HFi-GAN (\emph{HG V1}, \emph{HG V3}), Griffin-Lim (\emph{GL}) and the original ground-truth (\emph{GT}) waveform.

45 listeners were recruited using Prolific\footnote{\url{https://www.prolific.co/}}. A Qualtrics survey was generated using the \emph{Qualtreats} tool\footnote{\url{https://github.com/CSTR-Edinburgh/qualtreats}} which comprise 60 MUSHRA screens, split evenly into 3 tests, with each test assigned 15 listeners. Each screen presented the reference. Participants were then instructed to find the reference within those seven samples and assign it a score of 100, whilst rating all samples. 

\begin{figure}[!t]
    \centering % L B R T
    \includegraphics[clip, trim=0.5cm 0.5cm 0.5cm 1.2cm, width=\linewidth]{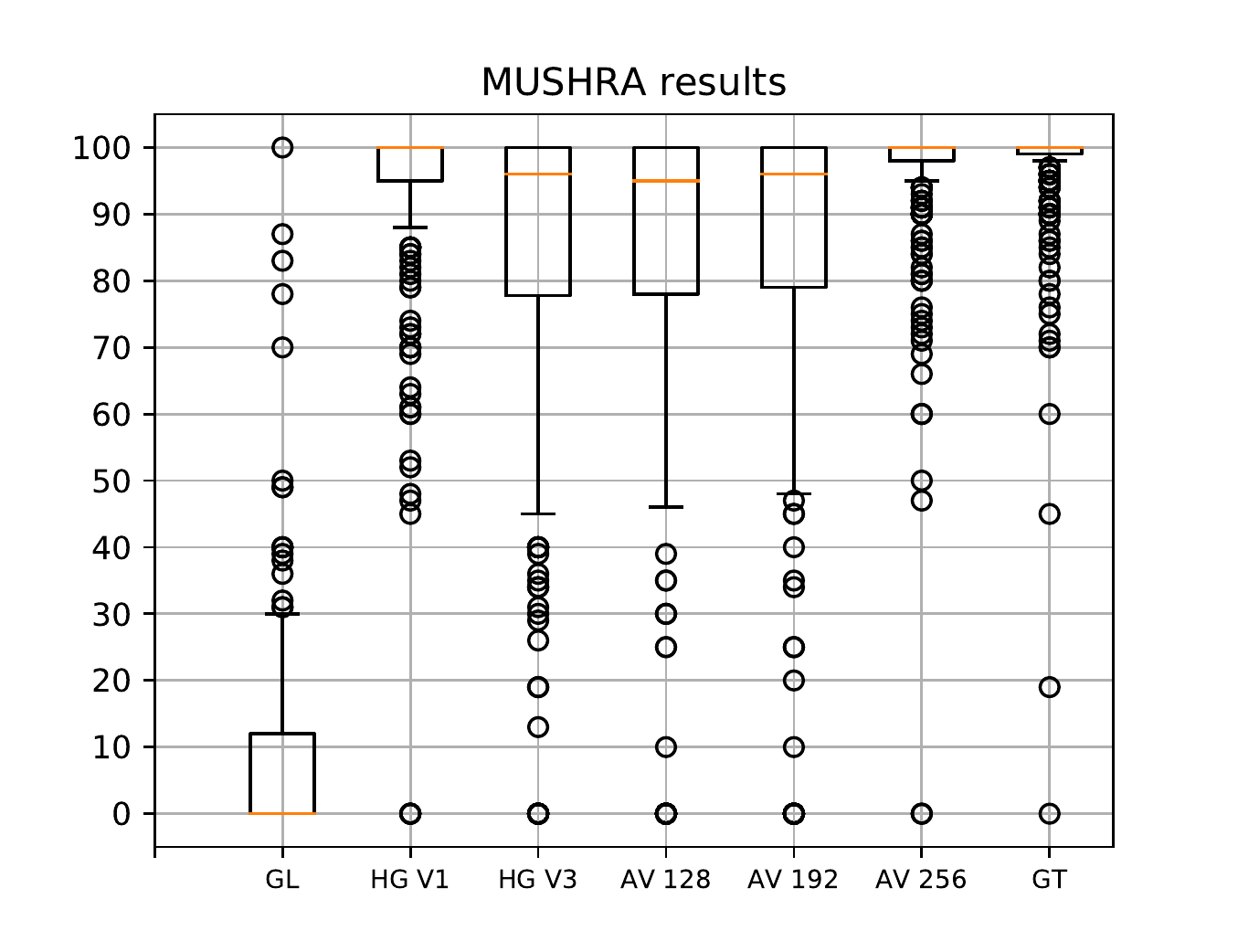}
    \caption{Results of the MUSHRA evaluation.}
    \label{fig:mushra}
\end{figure}

In line with the MUSHRA specification, participants who rated GT less than 90 in more than 15\% screens were excluded. This left 17 respondents and 340 MUSHRA screens for analysis. This large number of exclusions was a result of the high difficulty of distinguishing between GT, AV 256 and HG V1. Analysis without these exclusions yielded broadly similar results. Figure \ref{fig:mushra} shows the results. A pairwise $t$-test was used to evaluate statistical significance. Analysis yielded three sets of systems, \{GL\}, \{HG V3, AV 128, AV 192\} and \{HG V1, AV 256, GT\}. With $p > 0.01$ all systems were significantly different from those not in their own set, and there were no significant differences within sets.

%whose members were significantly different from all those in other sets at $p > 0.01$

Samples from all systems are available online\footnote{\url{https://jacobjwebber.github.io/autovocoder}}.

\subsection{Computational cost}

The computational performance of the \autovocoder decoder was compared with several waveform generation systems. Each system was timed generating individual utterances (with no batching) on a CPU. Such utterance-by-utterance processing represents many typical use cases, such as on-device synthesis. The systems chosen were Griffin-Lim and HiFi-GAN as used in the listening test, plus LPCNet and WaveRNN\footnote{\url{https://github.com/fatchord/WaveRNN}}.  A C implementation of LPCNet was used\footnote{\url{https://github.com/xiph/LPCNet}}, which may give it a performance advantage over the Python/PyTorch implementation of \autovocoder.

Timings were measured on a high-end 10 core desktop workstation processor (Intel i9-10850K). The value given in Table \ref{tab:speeds} was calculated by dividing the total duration of the generated audio by the total time taken to generate it. The measurement was for generating the test set of ten samples ten times. An average was taken of three runs. The test set contained a range of waveform duration from \SI{2.0}{\second} to \SI{9.7}{\second}.

LPCNet was not able to exploit the parallelism of the processor, and is therefore likely to perform relatively better on a system with fewer available cores. Time spent loading models from the file system was not found to be significant for any of the systems under test.

As shown in Table \ref{tab:speeds}, \autovocoder generates a waveform many times faster than the other waveform generators. Performance improvements of \autovocoder over autoregressive systems should be even larger on hyper-parallel architectures such as GPUs.

\begin{table}[!t]
    \caption{Speed of each system. Higher values indicate faster generation; a value below 1 indicates slower than real time.}

    \centering
    \begin{tabular}{c|c}
    System  &  \multicolumn{1}{p{3cm}}{Real-time factor}\\
    \hline
    Griffin-Lim     &   18.43 \\ 
    \Autovocoder 256,192,128 & 102.01, 101.34, 101.08 \\
    HiFi-GAN \emph{V1, V3} & 6.76, 42.18 \\
    WaveRNN & 0.47 \\
    LPCNet & 1.50 
    \end{tabular}
    \label{tab:speeds}
\end{table}

\section{Future application to TTS}

There are two primary ways that \autovocoder could be incorporated into a text-to-speech system. The first is by using the decoder architecture as the waveform generator within an end-to-end system. The second is by using \autovocoder to extract representations from waveforms, and using a TTS acoustic model to generate such representations instead of mel-spectrograms. The waveform could then be generated very quickly by the \autovocoder decoder.

% In the former approach the autoencoder task is purely for system design, as the autoencoder task is much faster to iterate with than a full end-to-end TTS system. 

% We took the former approach and embedded the autovocoder decoder within VITS \cite{vits}, replacing the standard HiFi-GAN-style decoder. Although a full description and evaluation of this system are left as future work, we include samples to show the viability of our approach as applied to TTS. 

\section{Conclusion}
\vspace{-0ex}
We propose an alternative to typical neural vocoders. \Autovocoder generates high-quality audio very quickly: it can reach the same quality as HiFi-GAN in fewer training updates, and is up to 16 times faster during generation. We intend that this system be used for the typical tasks of a neural vocoder, such as speech coding and speech synthesis. Verifying that speech synthesizers can accurately generate \autovocoder features remains as important future work.

\vspace{10pt}

%\section{Acknowledgements}
\noindent
{\bf Acknowledgements} Thanks are extended to Erfan Loweimi for his assistance with this work.

%\todo{Do listening test and analyse results, do speed test and put results in table for WaveRNN and LPCNet. Write about future work and generative models.}
% \bibliographystyle{IEEEbib}
\newpage
% \bibliography{mybib}

\section{References}
\small
\printbibliography[heading=none]

% \begin{thebibliography}{9}
% \bibitem[1]{Davis80-COP}
%   S.\ B.\ Davis and P.\ Mermelstein,
%   ``Comparison of parametric representation for monosyllabic word recognition in continuously spoken sentences,''
%   \textit{IEEE Transactions on Acoustics, Speech and Signal Processing}, vol.~28, no.~4, pp.~357--366, 1980.
% \bibitem[2]{Rabiner89-ATO}
%   L.\ R.\ Rabiner,
%   ``A tutorial on hidden Markov models and selected applications in speech recognition,''
%   \textit{Proceedings of the IEEE}, vol.~77, no.~2, pp.~257-286, 1989.
% \bibitem[3]{Hastie09-TEO}
%   T.\ Hastie, R.\ Tibshirani, and J.\ Friedman,
%   \textit{The Elements of Statistical Learning -- Data Mining, Inference, and Prediction}.
%   New York: Springer, 2009.
% \bibitem[4]{YourName17-XXX}
%   F.\ Lastname1, F.\ Lastname2, and F.\ Lastname3,
%   ``Title of your INTERSPEECH 2021 publication,''
%   in \textit{Interspeech 2021 -- 20\textsuperscript{th} Annual Conference of the International Speech Communication Association, September 15-19, Graz, Austria, Proceedings, Proceedings}, 2020, pp.~100--104.
% \end{thebibliography}

\end{document}